\documentclass[preprint,aps,showpacs]{revtex4}
\usepackage{epsfig}
\begin{document}

\title {Study on $N\bar{N}$ $S$-wave Elastic Cross Section and Possible Bound States
Within a Constituent Quark Model}

\author{Hourong Pang}
\affiliation{Department of Physics, Southeast University, Nanjing,
210094, P.R.China}

\author{Jialun Ping}
\affiliation{Department of Physics, Nanjing Normal University,
Nanjing, 210097, P.R.China}

\author{Fan Wang}
\affiliation{Joint Center of Particle, Nuclear Physics and
Cosmology, and Department of physics, Nanjing University, Nanjing,
210093, P.R.China}

\begin{abstract}
In the framework of a chiral constituent quark model, considering
the contributions of $\pi$ annihilation and one-gluon annihilation,
the proton-antiproton $S$-wave elastic scattering cross section
experimental data can be reproduced by adjusting properly one-gluon
annihilation coupling constant. Meanwhile, using the fixed model
parameter, we do a dynamical calculation for all possible $S$-wave
nucleon-antinucleon states, the results show that, there is no
$S$-wave bound state as indicated by a strong enhancement at
threshold of $p\bar{p}$ in $J/\psi$ and $B$ decays.

\end{abstract}


%\keywords{elastic cross section, $p\bar{p}$ threshold enhancement,
%constituent quark model}

\maketitle

\section{introduction}
Theoretical studies of baryon-antibaryon bound states date back to
the proposal of Fermi and Yang \cite{fermi} to make the pion with a
nucleon-antinucleon pair. The traditional $N\bar{N}$ interaction
studies, such as boundary condition model \cite{lomon}, optical
model \cite{cbdover} and coupled-channel models \cite{liu},
emphasized on the handling of the short-range part of $N\bar{N}$
interaction, and for the long range part they are quite similar. The
extensive and excellent reviews are given in Ref.\cite{richard}.
Although the agreement with scattering experimental data is
obtained, from the quantum chromodynamics (QCD) point of view it is
hard to image that a hadronic picture can be applied to such a
short-range where hadrons are ``overlapped" and the internal
structure of hadrons must be considered.

A possible way out is to start from QCD, the fundamental strong
interaction theory. Namely we should start with quark-gluon degree
of freedom, rather than the meson-baryon picture. QCD has been
already proved to be the right theory at high energy. At low energy,
because of the non-perturbative nature of QCD, one has to rely on
effective theories and/or QCD-inspired models to get some insight
into the phenomena of the hadronic world. The constituent quark
model \cite{constituent1,constituent2,constituent3,constituent4} is
one of them. It has been successful in describing hadron spectrum,
the baryon-baryon interactions and the bound state of two baryons,
the deuteron. Therefore, extending the constituent quark model to
$N\bar{N}$ study is an interesting practice.

The early studies in the traditional meson exchange framework found
that, if neglecting annihilation channels, many bound states might
be formed, while annihilation effects were included, the binding
force decreased and some bound states were washed out\cite{dover}.
Therefore, how to take into account the effect of annihilation in an
unified framework is important. We attempt to include the
contributions of the annihilations of a $q\bar{q}$ pair into meson
or into gluon in an unified manner, besides including $\pi$,
$\sigma$ and gluon exchange in the constituent quark model. In order
to keep the model well-describing baryon spectrum and $NN$
scattering data, all of the model parameters have been fixed as much
possible as those fixed in $NN$ interaction and baryon spectrum.

Perturbative QCD calculation showed that the gluon running coupling
constant decreasing with the increasing of momentum transfer. Here
we vary gluon annihilation coupling constant $\alpha'_s$ to see if
the proton-antiproton $S$-wave elastic cross section experimental
data\cite{experiment} can be well reproduced. For comparison, the
case of ignoring the contributions of $\pi$ annihilation is also
computed.

The BES collaboration in the radiative decay $J/\Psi \rightarrow
\gamma p\bar{p}$ observed a sharp enhancement at threshold in the
$p\bar{p}$ invariant mass spectrum \cite{bes}. They tried to fit the
enhancement by means of a $S$-wave Breit-Wigner resonance, and
obtained the resultant mass peak below threshold. Belle also
reported they observed an enhancement in the $p\bar{p}$ invariant
mass distribution near the threshold in the decays $B^+\rightarrow
K^+p\bar{p}$ and $B^0\rightarrow D^0p\bar{p}$ \cite{Belle}. Many
interpretations \cite{rosner,zou,datta,he} on the observation were
suggested. Here we apply the constituent quark model constrained by
baryon spectrum and $NN$ interaction and in addition considering the
contribution of annihilation fixed by proton-antiproton $S$-wave
elastic scattering cross section, to do a dynamical calculation for
all possible $S$-wave nucleon-antinucleon system to study if there
is $p\bar{p}$ $S$-wave bound state.

The paper is organized as follows. Sec. II explains the model
Hamiltonian, its parameters and the calculation method. Sec. III is
the results and discussions.

\section{Hamiltonian, model parameters and calculation method}

We take the chiral quark model used in the study of multi-quark
system, which essentially is an effective theory on exchanges of
Goldstone boson, scalar meson $\sigma$, as well as gluon between
quarks, and extend it to include the antiquarks and the annihilation
interaction, to study the nucleon-antinucleon system.

As the first step, here only $S$-wave states of the
nucleon-antinucleon pair are considered, i.e., the total orbital
angular momentum $L=0$, and we have $J=S$ (the total angular
momentum comes from quark spin only).

We start from a Hamiltonian which was used by Salamanca group for NN
interaction\cite{constituent3}. It is one of the chiral quark model
with only $\pi$ and $\sigma$ mesons. The quark-meson coupling
constant $\alpha_{ch}$ is fixed by $g^2_{NN\pi}/4\pi$
\[
\frac{g^2_{ch}}{4\pi}=(\frac{3}{5})^2
\frac{g^2_{NN\pi}}{4\pi}\frac{m^2_q}{M^2_N},~~~~\alpha_{ch}=\frac{g^2_{ch}}{4\pi}
\frac{m^2_\pi}{4m^2_u}.
\]
The quark mass $m_q$ and $\sigma$ meson
mass $m_\sigma$ are taken to be 313 MeV and 3.421 fm$^{-1}$,
respectively. Quark-gluon coupling constant $\alpha_s$ is determined
by the $\Delta-N$ mass difference. The confinement strength, $a_c$
is obtained from nucleon stability condition. Parameters and the
calculated deuteron properties are listed in Table I.
\begin{center}
Table I. Model Parameters, deuteron properties. $\delta_{OGE}$,
$\delta_{OPE}$ (MeV) are the contributions of gluon, $\pi$ exchanges
to $\Delta-N$ mass difference, respectively. \\
\begin{small}
\begin{tabular}{cccccc|ccccc} \hline \hline
$b$(fm) & $\alpha_s$ & $\alpha_{ch}$ & $\Lambda$(fm$^{-1})$ &
$a_c$(MeV fm$^{-2})$ & $V_0$(MeV)
&$\delta_{OGE}$&$\delta_{OPE}$&$B_{D}$(MeV)&$\sqrt(r^2)$(fm)&$P_D$
\\ \hline
0.518&0.485&0.027&4.2&46.938&-487.29 &145.6&148&2.0&1.96&4.86
\\
\hline\hline
\end{tabular}
\end{small}

\end{center}

To extend model from $NN$ systems to $N\bar{N}$ systems, we have to
take into account the annihilation contributions in addition to the
scattering ones: the relevant scattering and annihilation Feynman
diagrams are shown in Fig.1 and Fig.2.

\begin{figure}[!tbh]
\setlength{\abovecaptionskip}{0pt}
\includegraphics[height=2.2in]{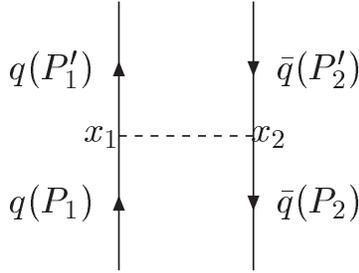} \caption{The exchange
diagram between quark and antiquark} \label{Fig1}
\end{figure}

\begin{figure}[!tbh]
\setlength{\abovecaptionskip}{0pt}
\includegraphics[height=2.2in]{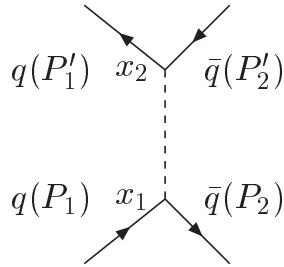} \caption{The annihilation
diagram between quark and antiquark} \label{Fig2}
\end{figure}

Taking $\pi$ exchange as an example (in orbit and spin space), the
T-matrix of $\pi$ exchange diagram between quark and antiquark can
be written as
\begin{eqnarray}
&&\bar u(p'_1,s'_1) \gamma^5 u(p_1,s_1) \frac{-F^2(\vec
q^2)}{q^2+m^2_{\pi}}g^2_{ch} \bar v(p_2,s_2) \gamma^5 v(p'_2,s'_2),
\end{eqnarray}
 here, a form factor $F(\vec q^2)$ is assumed to be $(\Lambda^2
/(\Lambda^2 +\vec q^2))^{1/2}$. $p_1,s_1,p'_1,s'_1$ are the
four-vector momenta and spin z-projections of initial quark and
final quark, respectively. $p_2,s_2,p'_2,s'_2$ are the four-vector
momenta and spin $z$-projections of initial antiquark and final
antiquark, respectively. $u$ and $v$ are assumed to be free Dirac
spinors for quarks and antiquarks, respectively, $m_{\pi}$ is the
$\pi$ mass.

Similarly, T-matrix of $\pi$ annihilation diagram can be written as
\begin{eqnarray}
&&-\bar u(p'_1,s'_1) \gamma^5 v(p'_2,s'_2)
\frac{-1}{q^2+m^2_{\pi}}g^2_{ch} \bar v(p_2,s_2) \gamma^5 u(p_1,s_1)
\end{eqnarray}
using Fierz identities for Dirac matrices, a four-fermion matrix can
be expressed as a linear superposition of other matrices with a
changed sequence of spinors \cite{zong}.

\[ ({\bar a}O_i b)({\bar c} O^i d)=\sum^{16}_{k=1}({\bar a}O_k d)({\bar c}
O^k b) \]

Then we take the non-relativistic limit and transform the potential
to space-time representation, we have
\begin{eqnarray}
V^{{\pi}_{a}}_{q \bar q}&=&4\pi
\frac{g^2_{ch}}{4\pi}\frac{1}{m_{\pi}^2-(m_1+m_2)^2}
\delta(r)(-\frac12-\frac12\sigma_i\cdot
\sigma_j)(\frac32-\frac12\tau_i\cdot \tau_j)
(\frac13+\frac12\lambda_i\cdot \lambda_j)~~~~~~  \\
V^{\pi_{e}}_{q\bar q}(r) &=& -\frac{1}{12}\frac{g^2_{ch}}{4\pi}
\frac{m^2_\pi}{m_{q_i}m_{q_j}}\left[\frac{e^{-m_\pi
r}}{r}-\frac{\Lambda^2}{m^2_\pi} \frac{e^{-\Lambda r}}{r}\right]
\frac{\Lambda^2}{\Lambda^2-m_\pi^2} \sigma_i \cdot \sigma_j \tau_i
\cdot  \tau_j \\
V^{g_{a}}_{q \bar q}&=& -\frac{4\pi
\alpha'_s}{(m_1+m_2)^2}\delta(r)\frac{1}{4}
                        (\frac{16}{9}-\frac13 \lambda^c_i\cdot \lambda^c_j)
                        (\frac12+\frac12 \tau_i\cdot \tau_j)
                        (-\frac32+\frac12 \sigma_i\cdot \sigma_j)
\end{eqnarray}
$V^{\pi_{e}}_{q\bar q}$ and $V^{\pi_{a}}_{q\bar q}$ are the
effective potential from $\pi$ exchanges and annihilation between
quark and antiquark, respectively. $V^{g_{a}}_{q\bar q}$ are the
effective potential from one gluon annihilation. Because here $N$
and $\bar{N}$ are both color singlet and there is no quark exchange
between $N$ and $\bar{N}$, one gluon exchange between $N$ and
$\bar{N}$ does not contribution at all.

Since we are taking into account the contributions of the lowest
order to the $S$-wave at this step, under static approximation the
contribution from $\sigma$ meson annihilation in the present case
vanishes. The $\sigma$ exchange potential between quark and
antiquark in the non-relativistic limit in coordinate space can been
written as
\begin{eqnarray}
 V^{\sigma_{e}}_{q\bar q}(r) &=&
-\frac{g^2_{ch}}{4\pi}\frac{\Lambda^2}{\Lambda^2 - m_{\sigma}^2}
\left[\frac{e^{-m_\sigma r}}{r}-\frac{e^{-\Lambda r}}{r}\right].
\end{eqnarray}
The detailed derivation of these effective potentials between quark
and antiquark can be found in Ref.\cite{chang}.

Therefore, the Hamiltonian of $N \bar N $ system is
\begin{eqnarray}
H_{p\bar p}  &=&  \sum_{i=1}^6 (m_i+\frac{p_i^2}{2m_i})-T_{CM}
+\sum_{i>j=1}^6
    [ V_{conf}(r_{ij})
+ V^{e}(r_{ij}) +V^{a}_{q \bar q}(r_{ij})]   \nonumber \\
 V^{e}(r_{ij}) &=&V^{\pi_{e}}_{qq,q \bar q}(r_{ij})
                   +V^{\sigma_{e}}_{qq,q \bar q}(r_{ij})
                   +V^{gluon_{e}}_{qq,q \bar q}(r_{ij}) \\
 V^{a}_{q \bar q}(r_{ij})&=&V^{\pi_{a}}_{q \bar q}(r_{ij})
     +V^{\sigma_{a}}_{q \bar q}(r_{ij})
     +V^{g_{a}}_{q \bar q}(r_{ij})\nonumber \\
V_{conf}(r_{ij})&=& -a_c\lambda_i^c \cdot\lambda_j^c r_{ij}+V_0.
\nonumber
\end{eqnarray}
We should note that, in order to keep the model well-describing
baryon spectrum and $NN$ scattering data, all of the model
parameters related to $NN$ system are unchanged, only one parameter
$\alpha'_s$ connected with annihilation has been left to be
adjusted.

We use Kohn-Hulthen-Kato variational method for bound and scattering
problems in an unified framework\cite{khk,oka}. \\
(i) for bound problem

Following the cluster model approach, the resonating group method
(RGM) wave function is written as
\[\Psi_{LM}(\xi_{B_1},\xi_{B_2},\vec R)={\cal A}[\phi_{B_1}(\xi_{B_1})\phi_{B_2}(\xi_{B_2})\chi(\vec R)] \]
here $\vec R$ is the relative coordinate between the clusters of
${B_1}$ and ${B_2}$. $\cal A$ is the antisymmetrization operator but
in fact there is no need for this antisymmetrization for $N\bar{N}$
system because $q$ and $\bar{q}$ can be treated as different
particles. $\phi_{B_1}$ and $\phi_{B_2}$ are the internal wave
functions of two quark clusters, $\chi( \vec R)$ is relative motion
wave function.

The relative motion wave function is expanded into partial waves
\begin{eqnarray}
\chi(\vec R)&=&\sum_L \chi^L(R)Y^{LM}(\hat R) = \sum_L \sum_{i=1}^N
c_i^L \chi_i^L(R,S_i)Y^{LM}(\hat R)
\end{eqnarray}
 with
\begin{eqnarray}
 \chi_i^L(R,S_i)Y^{LM}(\hat R)&=&(\frac{3}{2\pi b^2})^{3/4}\int
exp({-\frac{3}{4b^2}(\vec R-\vec S_i)^2})Y^{*LM}(\hat
R)d\Omega_{R}Y^{LM}(\hat S_i)d\Omega_{S_i} \nonumber \\
 & =&4\pi (\frac{3}{2\pi
b^2})^{3/4} exp({-\frac{3}{4b^2}(R^2+S_i^2)})i_L(\frac{3}{2b^2}RS_i)
Y^{LM}(\hat R)
\end{eqnarray}
 where $i_L$ is the modified spherical Bessel
function. In this paper, only $S$ wave ($L=0$) is taken into
account.

Adding the center of mass motion, the wave function of six quarks
can be written as the production of the single-particle orbital wave
function with different reference centers, i.e.,
\begin{eqnarray}
 \Psi_{6q} & =& {\cal
A}  \sum_{L_k} \sum_{i=1}^{n} C_{{L_k},i}
  \int d\Omega_{S_i}
  \prod_{\alpha=1}^{3} \psi_\alpha(\vec{S_i})
  \prod_{\beta=4}^{6} \psi_\beta(-\vec{S_i})
    [[\Phi_p^{j_1f_1}\Phi_{\bar p}^{j_2f_2}]^{I,S}Y^L(\hat S_i)]^J
   [\chi_c(B_1)\chi_c(B_2)]^{[\sigma]}~~~~~~
\end{eqnarray}
here $\psi_\alpha(\vec{S_i})$ and $\psi_\beta(-\vec{S_i})$ are the
single-particle orbital wave function with different reference
centers, $L_k$ is the coupled channels index.

Via the variation with respect to the relative motion wave function
$\chi(R)$, the RGM equation
\begin{eqnarray}
\int H(\vec R,\vec R')\chi (\vec R') d \vec R'&=&E\int N(\vec R,\vec
R')\chi (\vec R') d \vec R'
\end{eqnarray}
 becomes an algebraic eigenvalue
equation
\begin{eqnarray}
 \sum_{j,{L_k'}}C_{j,{L_k'}}H^{L_k,L_k'}_{i,j}&=&E\sum_j
C_{j,L_k}N^{L_k}_{i,j}.
\end{eqnarray}
 where $N^{L_k}_{i,j}$ and
$H^{L_k,L_k'}_{i,j}$ are the wave function overlaps and Hamiltonian
matrix elements, respectively.\\
(ii) for scattering problem

The wave function of the relative motion is expanded by
\begin{eqnarray}
\chi(\vec R)&=&\sum_L \chi^L(R)Y^{LM}(\hat R) = \sum_L \sum_{i=0}^n
c_i^L \tilde\chi_i^L(R,S_i)Y^{LM}(\hat R)
\end{eqnarray}
 here
\begin{eqnarray}
 \tilde\chi_i^L(R,S_i)&=&\left \{  \begin{array}{cc}  \alpha_i \chi_i^L(R,S_i)&R < R_c \\
                                                                ( h_L^{(-)}(kR)-s_i h_L^{(+)}(kR))kR & R> R_c \\
                                      \end{array}
                            \right.
\end{eqnarray}
with
\begin{eqnarray}
\chi_i^L(R,S_i)=\frac{u_i^L(R,S_i)}{R}&=&4\pi (\frac{3}{2\pi
b^2})^{3/4} exp[{-\frac{3}{4b^2}(R^2+S_i^2)}]i_L(\frac{3}{2b^2}RS_i)
\end{eqnarray}
 where
$i_L$ is the modified spherical Bessel function. $h_L^{(\pm)}$ is
the $L$-th spherical Hankel function, $k=\sqrt(2\mu E_{rel}) $.

The constants $\alpha_i$ and $s_i$ are determined by the condition
that the relative motion wave function for $R<R_c$ and $R>R_c$
smoothly connect at $R=R_c$.

Using the normalization condition $c_0=1-\sum_{i=1}^n c_i$ and
varying with respect to parameters $c_i~(i=1,...,n)$, we have
\[  ([u_i-u_0]\L_L u_t)=0 ,  i=1,...,n\]
with  a symbol $(f \L_L g)$ by
\[
(f \L_L
g)=\int_0^{\infty} \int_0^{\infty} f(R) \L_L(R,R') g(R') dRdR'\],
\[ \L_L (R,R')=\int [\phi^{+}_A(\xi_A)\phi^{+}_B (\xi_B)u_i^L(R)Y^*_{LM}(\hat R)]
(H-E){\cal A}[\phi_A(\xi_A)\phi_B (\xi_B)u_j^L(R)Y_{LM}(\hat R)],
\]
then we have the n linear equations for the $c_i$'s,
\begin{eqnarray}
 &&\sum_{j=1}^n
\L_{ij} c_j =M_i
\end{eqnarray}
where
\begin{eqnarray}
 \L_{ij}&=&K_{ij}-K_{i0}-K_{0j}+K_{00}, M_i=K_{00}-K_{i0}
\end{eqnarray}
with
\begin{eqnarray}
 K_{ij}&=&(u_i \L_L u_j)
\end{eqnarray}
  The final phase shift is
\begin{eqnarray}
 s_{st}&=&s_t+i\alpha\sum_{i=0}^n K_{0i}c_i
\end{eqnarray}
here $\alpha=\mu/k $, $\mu$ is the reduced mass. The difference $s_t
-s_{st}$ is a good measure to check the accuracy of the calculation.
In order to calculation  $K_{ij}$ (term(18)) as analytically as
possible, there is a very useful skill mentioned in Ref.\cite{khk}.

\section{Results and Discussions}

There are four possible states for an S-wave nucleon-antinucleon
systems with different isospin and total angular momentum J
respectively. They are $IJ^{PC}=11^{--}$, $10^{-+}$, $01^{--}$,
$00^{-+}$. Since running coupling constant would change with
momentum transfer, here we adjust gluon annihilation coupling
constant $\alpha'_s$ from being smaller than exchange coupling
constant $\alpha_s$ to larger than $\alpha_s$.

In experiment one measures the spin-averaged scattering amplitudes,
and the elastic scattering amplitude is the sum of the isospin $I=0$
and $I=1$ amplitudes with equal weights. According these, we
calculated the total proton-antiproton elastic cross sections, and
compare it with experimental data.

We calculated $p \bar p$ $S$-wave elastic cross section including
$\pi $ annihilation and gluon annihilation with different gluon
annihilation coupling constant $\alpha'_s$, the results are shown in
Fig.3. We found that, if we let $\alpha'_s=0$ (i.e.,we did not take
into account the effect of gluon annihilation) the cross section
would be very larger than experimental data. It implies that, if
excluding the gluon annihilation process, $N \bar N$ system is more
attractive than $NN$ system and many bound states might be formed,
which is consistent with the results in Ref.\cite{dover,buck}. When
we add the contribution of gluon annihilation, with a gradual
increasing $\alpha'_s$, the cross section will decrease quickly,
especially in the low energy region. And the larger $\alpha'_s$ is,
the smoother the change of cross section with the scattering energy
is. If we choose $\alpha'_s$ equal to $\alpha_s$, the cross section
will be below the experimental data. When we choose $\alpha'_s$ is
about the one third of gluon exchange coupling constant $\alpha_s$
($\alpha'_s=\alpha_s/3$), the cross section became close to
experimental data, the difference between theoretical cross section
and experimental one would not more than 5 mb.
\begin{center}
%\begin{figure}[!tbh]
%\setlength{\abovecaptionskip}{0pt}
%\fbox{\includegraphics[height=3.5in]{Fig3.eps.EPS}}
%\caption{proton-antiproton elastic cross sections, including $\pi$
%annihilation and one-gluon annihilation with different coupling
%constants. The full squares show experimental data of total elastic
%cross section. The open squares are that of S-wave
%component\cite{experiment}.} \label{Fig3}
%\end{figure}
\epsfxsize=5.in \epsfbox{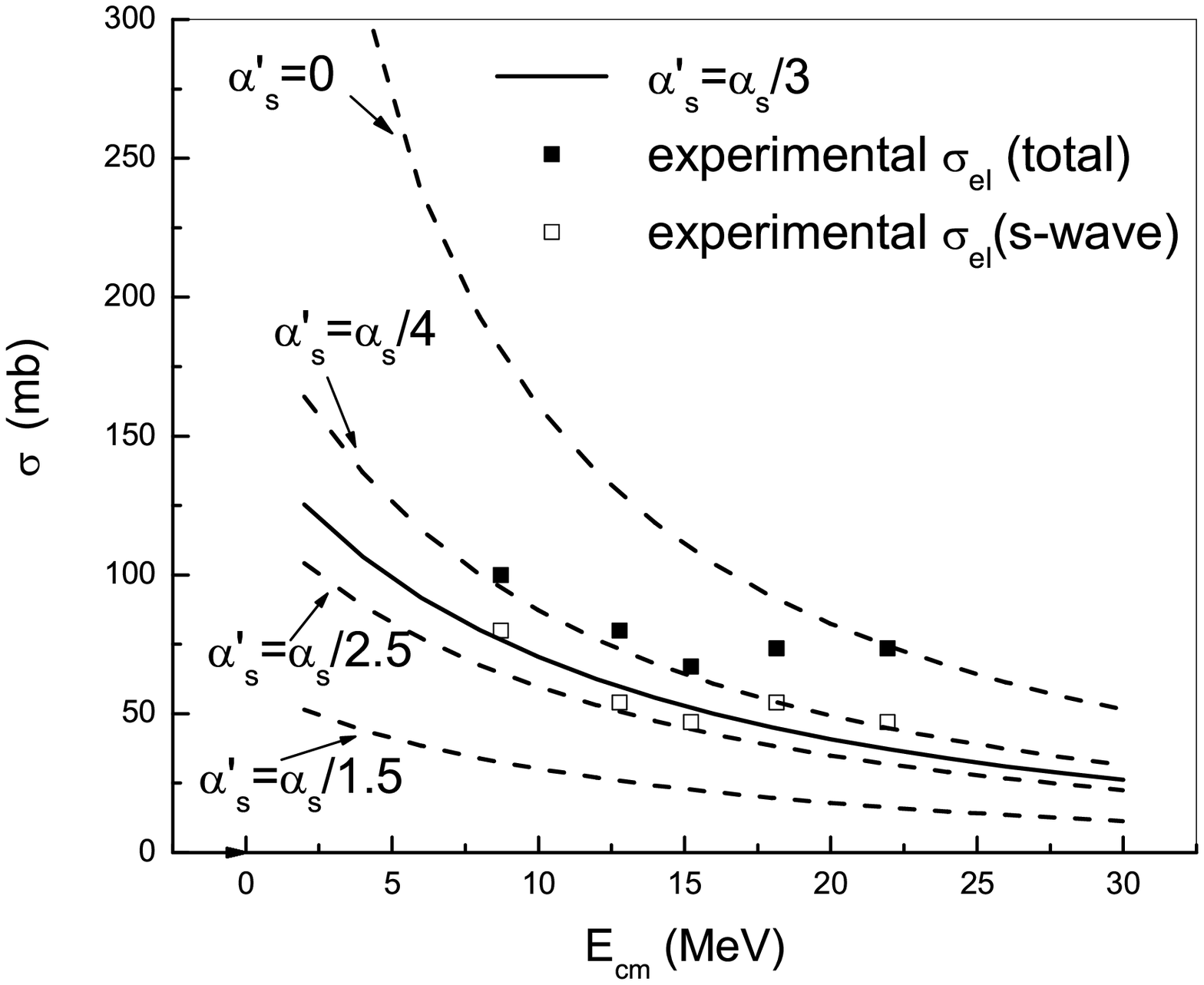}

Fig.3 proton-antiproton elastic cross sections, including $\pi$
annihilation and one-gluon annihilation with different coupling
constants. The full squares show experimental data of total elastic
cross section. The open squares are that of $S$-wave
component\cite{experiment}.

%\begin{figure}[!tbh]
%\setlength{\abovecaptionskip}{0pt}
%\fbox{\includegraphics[height=3.5in]{Fig4.EPS}} \caption{Same as
%Fig.3 but $\pi$ annihilation excluded.} \label{Fig4}
%\end{figure}
\epsfxsize=5.in \epsfbox{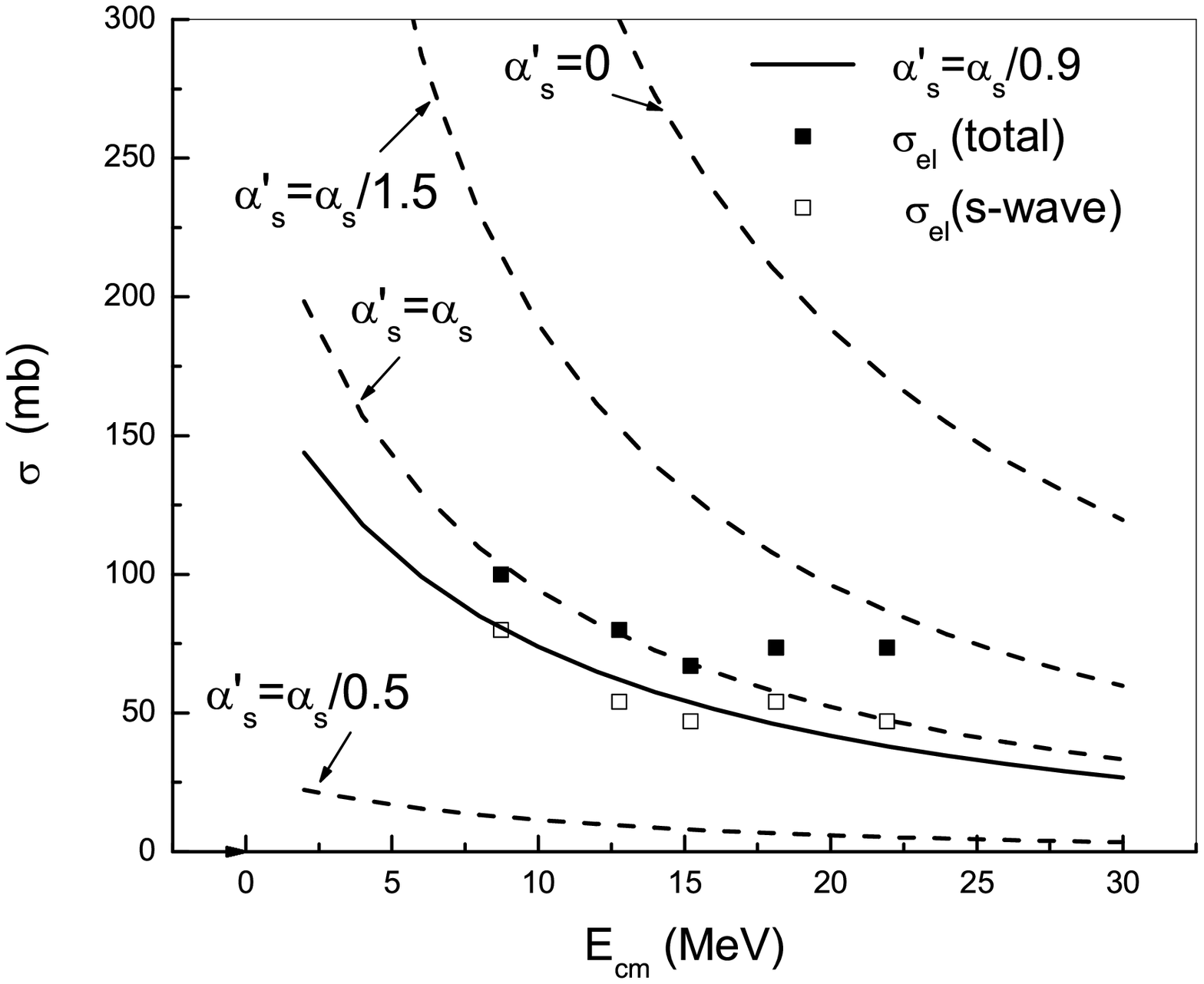}

Fig.4 Same as Fig.3 but $\pi$ annihilation excluded.
\end{center}

Fig.3 proton-antiproton elastic cross sections, including $\pi$
annihilation and one-gluon annihilation with different coupling
constants. The full squares show experimental data of total elastic
cross section. The open squares are that of $S$-wave component
\cite{experiment}. Since annihilation process would happen at short
range, and $\pi$ term might play an important role only at long
range, so we do a calculation by excluding the $\pi$ annihilation
term. Fig.4 shows the $p \bar p$ cross section with the different
$\alpha'_s$ in the case of excluding the contribution of $\pi$
annihilation. We found that, larger $\alpha'_s$ corresponds to
smaller cross section. In order to reproduce $p \bar p$ $S$-wave
elastic cross section, the effective gluon annihilation coupling
constant $\alpha'_s$ should increase to compensate for the absence
of $\pi$ annihilation. The proper value is
$\alpha'_s$=$\alpha_s/0.9$, which is very close to gluon exchange
coupling constant.

The above results implies that both gluon annihilation and $\pi$
annihilation provide effective repulsion, which would decrease the
$S$-wave cross section. By adjusting the strength of annihilation
term properly, proton-antiproton $S$-wave cross section experimental
data can be reproduced no matter the $\pi$ annihilation is included
or not.

Taking $\alpha'_s=\alpha_s/3$ and $\alpha'_s=\alpha_s/0.9$ as
examples, we give the contributions of different spin and isospin
components to the total cross section in Fig.5 and Fig.6 for
including and excluding $\pi$ annihilation, respectively. The
results show that, the contribution of $IJ=00$ channel is always
very small no matter the $\pi$ annihilation is included or not; the
$IJ=01$ channel contribution to cross section is larger than $IJ=10$
channel when $\pi$ annihilation is included; while the former
contribution is smaller than the latter if the $\pi$ annihilation is
excluded; and with the increasing of scattering energy the
difference between these two channels contributions are both
decreasing in these two cases; For the case of including $\pi$
annihilation, the total cross section is close to the $IJ=11$
channel ones when scattering energy $E_{cm}$ is below 15 MeV and it
will be close to $IJ=10$ channel ones when $E_{cm}$ is above 15 MeV.
If excluding the contribution of $\pi$ annihilation, the total cross
section is very close to $IJ=11$ channel in the whole energy range.

%\begin{figure}[!tbh]
%\setlength{\abovecaptionskip}{0pt}
%\fbox{\includegraphics[height=3.5in]{Fig5.EPS}} \caption{S-wave
%elastic cross sections with $\alpha'_s= \alpha_s/3.0$ in the case of
%including $\pi$ annihilation, for channel IJ=11,10,01,00,
%respectively. } \label{Fig5}
%\end{figure}

%\begin{figure}[!tbh]
%\setlength{\abovecaptionskip}{0pt}
%5\fbox{\includegraphics[height=3.5in]{Fig6.EPS}} \caption{Same as
%Fig.5 for the case of excluding $\pi$ annihilation with $\alpha'_s=
%\alpha_s/0.9$.} \label{Fig6}
%\end{figure}
\begin{center}
\epsfxsize=5.in \epsfbox{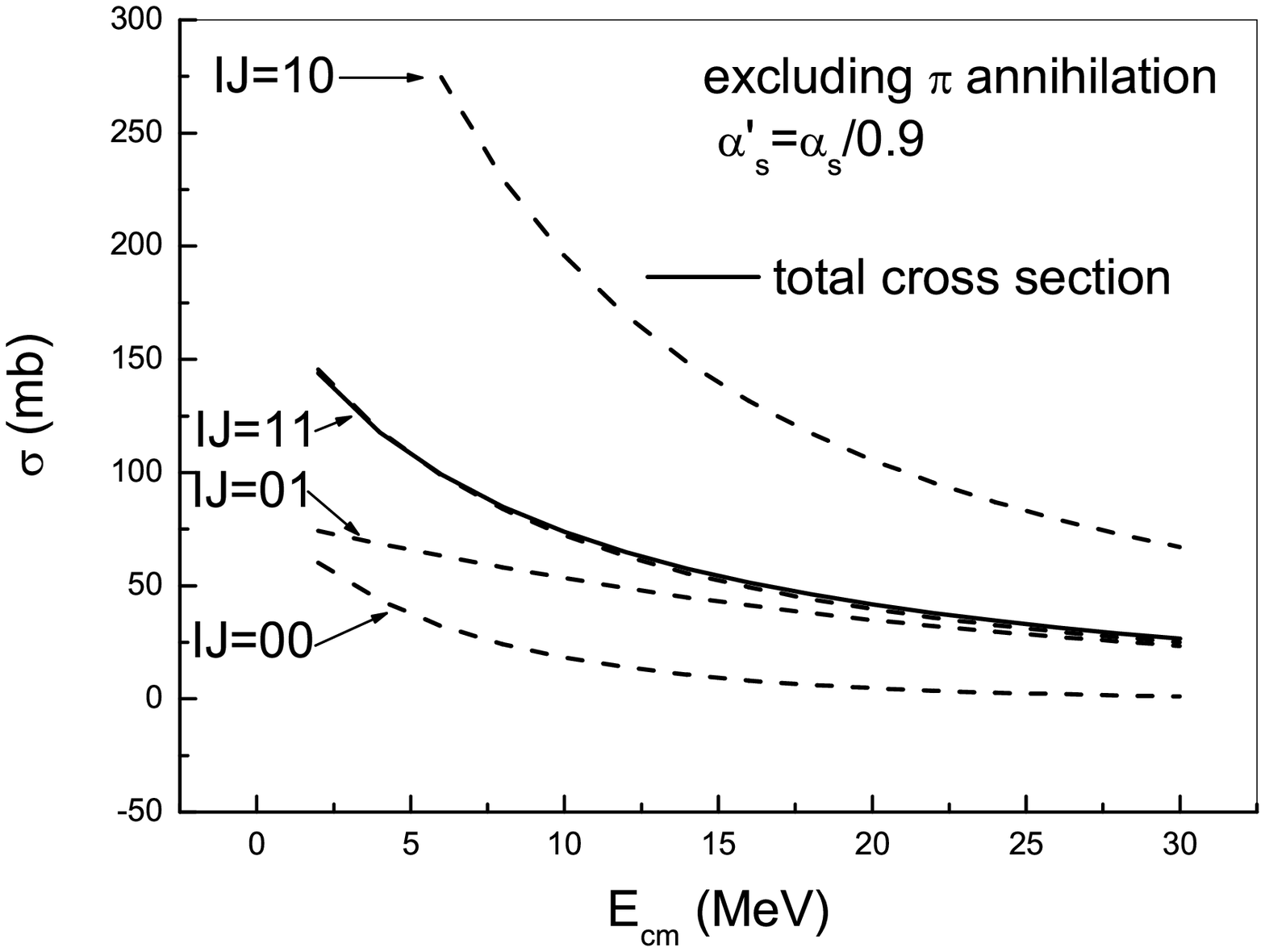}

Fig.5 $S$-wave elastic cross sections with $\alpha'_s= \alpha_s/3.0$
in the case of including $\pi$ annihilation, for channel
$IJ=11,10,01,00$, respectively.

\epsfxsize=5.in \epsfbox{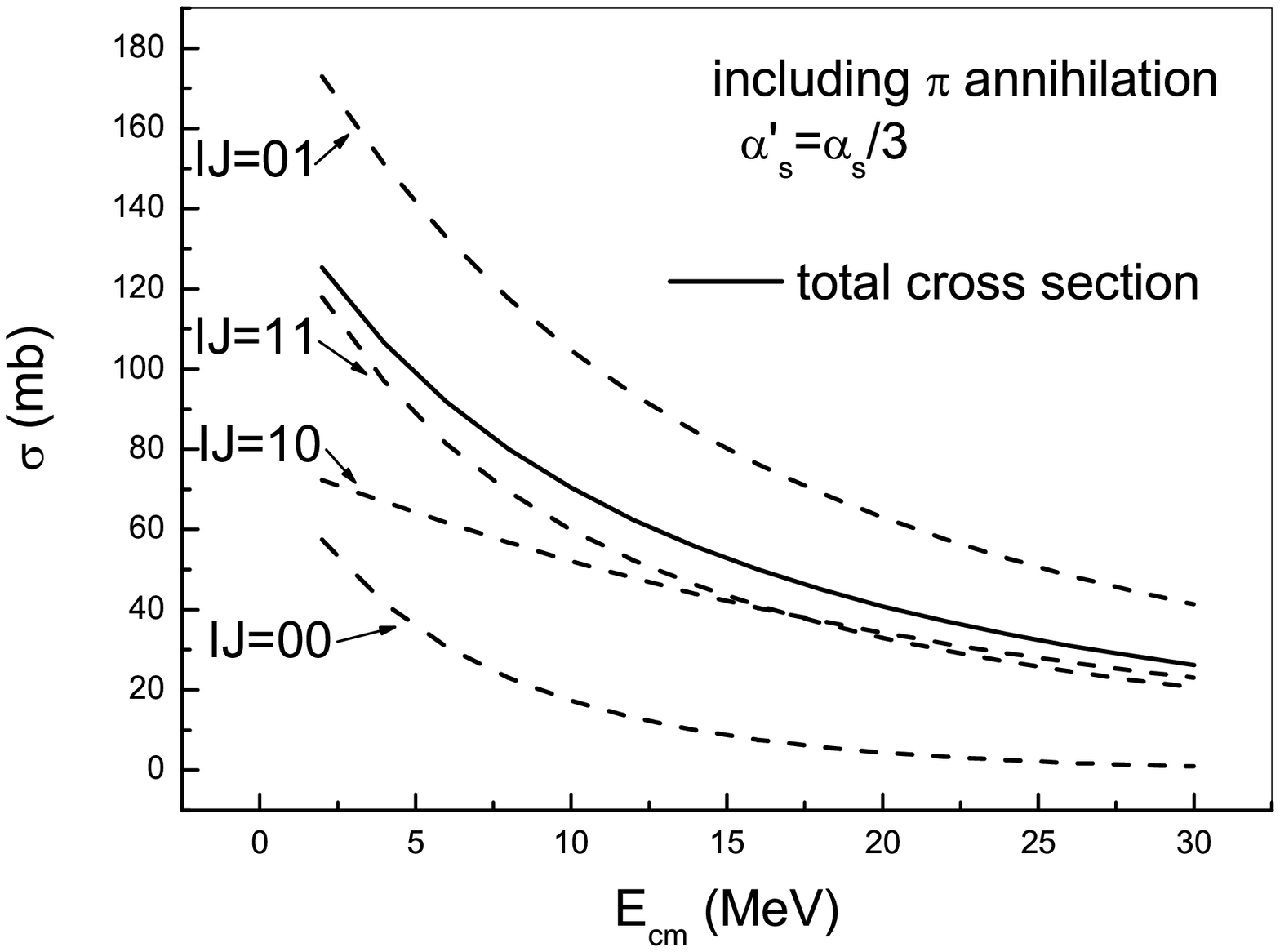}

Fig.6 Same as Fig.5 for the case of excluding $\pi$ annihilation
with $\alpha'_s= \alpha_s/0.9$.
\end{center}

The effective potentials of $IJ=11,10,01,00$ channels are given in
Fig.7 and Fig.8 corresponding to different annihilation coupling
constants. From these figures we found that $IJ=00$ channel has very
small intermediate attraction no matter what parameters are chosen,
which is consistent with the results of cross section in Fig.5-6.
The other three channels all have attraction to some extent.
Excluding the contribution of $\pi$ annihilation there will be more
attraction left in the other three channels, and the minimum values
of potential are all at the separation of ~1.0 fm.

The BES collaboration in the radiative decay $J/\psi \rightarrow
\gamma p \bar p$ observed a sharp enhancement at threshold in the $p
\bar p$ invariant mass spectrum\cite{bes}, and they obtained the
mass peak below the threshold of $p \bar p$. The above effective
potentials give qualitative information only, in order to see
whether there is a $p \bar p $ bound state, we do a dynamical
calculation. Taking into account the contribution of annihilation
and using the model parameters (in Fig.5-6) determined by
proton-antiproton S-wave elastic scattering cross section, we solve
RGM equation for all possible S-wave nucleon-antinucleon states, to
see if there is $p \bar p$ S-wave bound state. Here, we use 10 basis
wave functions to expand the wave function of the relative motion,
the boundary point is at 5.8 fm. We find that, although the four
possible channels with different isospin and spin quantum numbers
all have intermediate range attraction ( from several Mev to about
50 MeV), there is no bound state in dynamical calculation. Moreover,
there is no resonant state below the system mass of 3.9 GeV in our
calculation. That is to say, if we determine the model parameter of
quark-antiquark annihilation by fitting proton-antiproton S-wave
elastic cross section experimental data, our model does not support
a tight bound state claimed by BES and Belle experimental groups.

%\begin{figure}[!tbh]
%\setlength{\abovecaptionskip}{0pt}
%\fbox{\includegraphics[height=3.5in]{Fig7.EPS}}
%\caption{proton-antiproton effective potential with $\alpha'_s=
%\alpha_s/3.0$, including $\pi$ annihilation, for channel
%IJ=11,10,01,00, respectively. } \label{Fig7}
%\end{figure}

%\begin{figure}[!tbh]
%\setlength{\abovecaptionskip}{0pt}
%\fbox{\includegraphics[height=3.5in]{Fig8.EPS}} \caption{Same as
%Fig.7 for excluding $\pi$ annihilation with $\alpha'_s=
%\alpha_s/0.9$.} \label{Fig8}
%\end{figure}
\begin{center}
\epsfxsize=5.in \epsfbox{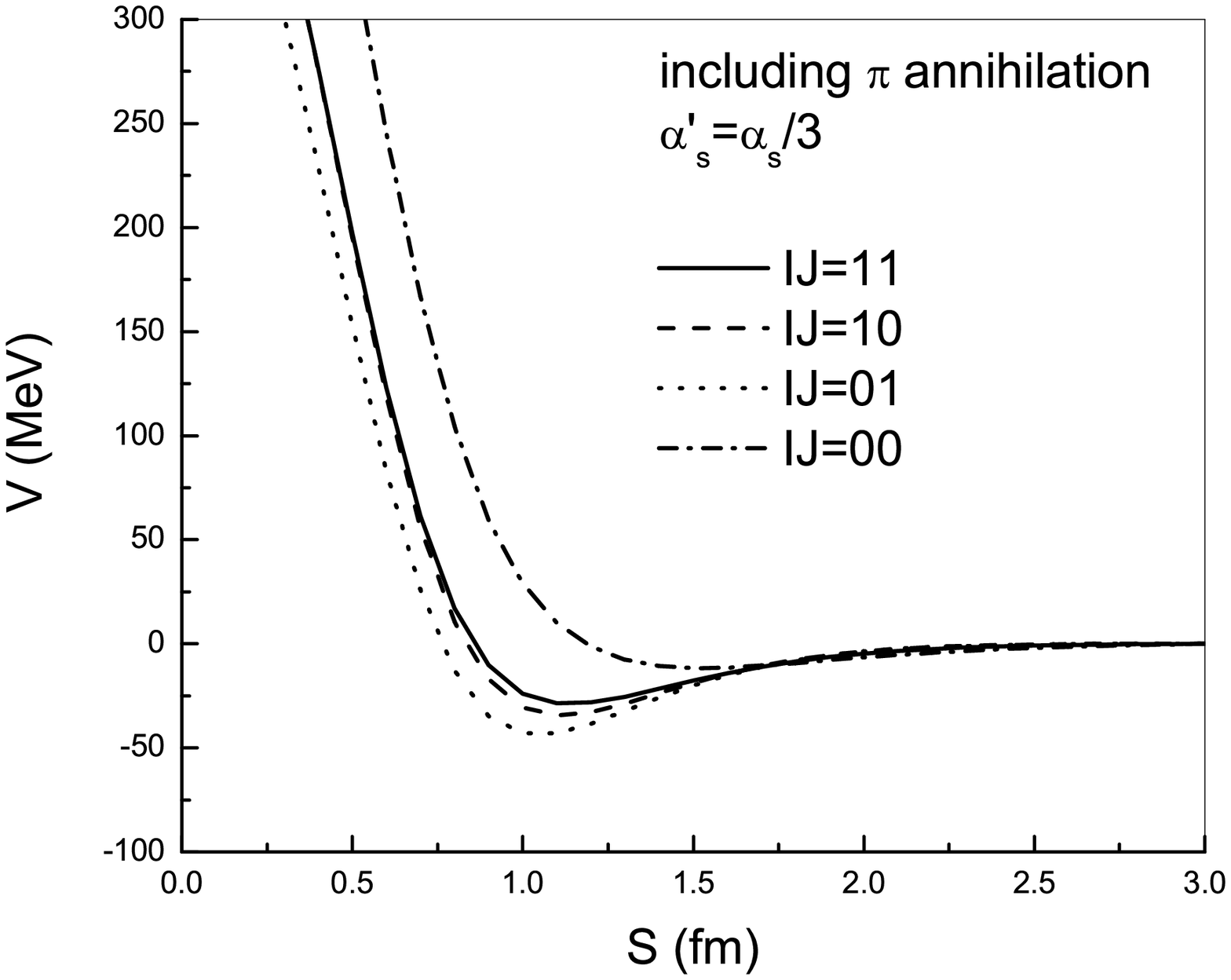}

Fig.7 Proton-antiproton effective potential with $\alpha'_s=
\alpha_s/3.0$, including $\pi$ annihilation, for channel
$IJ=11,10,01,00$, respectively.
\end{center}

To sum up, in the framework of chiral quark model, adding the
contribution of one gluon and $\pi$ annihilation, we can reproduce
proton-antiproton $S$-wave elastic cross section experimental data
by adjusting the coupling constant of gluon annihilation term. Using
the model parameters determined by $p\bar{p}$ scattering cross
section, our dynamical calculation for $N \bar N$ system with
$IJ=11,10,01,00$, quantum numbers does not find an $S$-wave bound
state.

\begin{center}
\epsfxsize=5.in \epsfbox{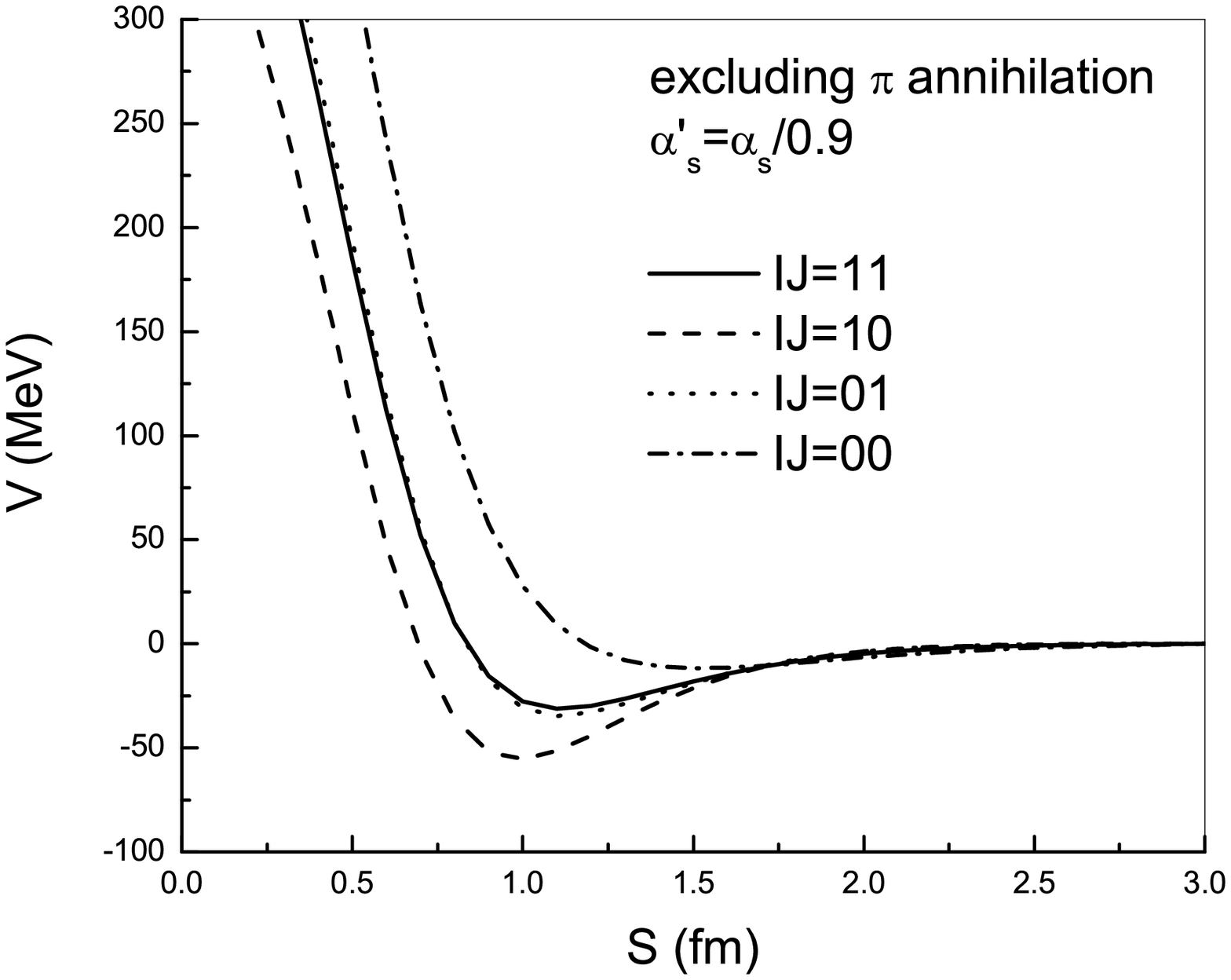}

Fig.8 Same as Fig.7 for excluding $\pi$ annihilation with
$\alpha'_s= \alpha_s/0.9$.
\end{center}

Obviously our conclusion of no $p\bar{p}$ bound state is based on
the assumption that the chiral constituent quark model is suitable
for $N\bar N$ system. Our conclusion is also based on the assumption
that the multi $\pi$ channel coupling effect, which is possible
within the chiral constituent quark model, can be neglected. In
addition all of the hidden color channels coupling effects have been
omitted. All of these effects should be studied further.

\section{Acknowledgments}

This work is supported by the NSFC under Contract No.10505006,
10375030 and 90503011. We would like to thank Dr. Chang Chao-hsi for
his collaboration in developing this model.

\end{document}